\documentclass[aps,prd,groupedaddress,showpacs]{revtex4}
\usepackage[parfill]{parskip}    
\usepackage{graphicx}
\usepackage{amssymb}
\usepackage{epstopdf}
\usepackage{color}
\usepackage{float}
\usepackage{ulem}
\usepackage{graphicx}
\usepackage{pdfpages}
\usepackage{amsmath}
\usepackage[colorlinks=true,citecolor=blue,urlcolor=magenta,breaklinks]{hyperref}
\usepackage{xcolor}     

\usepackage{amssymb}
\usepackage{graphicx,fancyhdr,url,amsfonts,amsthm}   
\usepackage{mathrsfs}
\usepackage{mathtools}
\usepackage{amsmath}
\usepackage{hyperref}
\usepackage{xcolor}

\newtheorem{thm}{Theorem}[section]
 
 \newtheorem{lem}[thm]{Lemma}

 \theoremstyle{definition}
 \newtheorem{defn}[thm]{Definition}
 \theoremstyle{remark}
 \newtheorem{rem}[thm]{Remark}
 
 \numberwithin{equation}{section}

 \newcommand{\dsp}{\displaystyle}

\newcommand{\bt}{\begin{theorem}}
\newcommand{\et}{\end{theorem}}
\newcommand{\be}{\begin{equation}}
\newcommand{\ee}{\end{equation}}
\newcommand{\beqn}{\begin{eqnarray}}
\newcommand{\eeqn}{\end{eqnarray}}
\newcommand{\bqa}{\begin{eqnarray*}}
\newcommand{\eqa}{\end{eqnarray*}}

\newcommand{\bproof}{\begin{proof}}
\newcommand{\eproof}{\end{proof}}
\newcommand{\bmat}{\left ( \begin{matrix}}
\newcommand{\emat}{\end{matrix} \right )}
\newcommand{\bmatn}{\begin{matrix}}
	\newcommand{\ematn}{\end{matrix} }

\begin{document}

\title{Fractional Einstein field equations in $2+1$ dimensional spacetime.}

\author{E. Contreras }
\email{ernesto.contreras@ua.es}
\affiliation{Departamento de F\'{\i}sica Aplicada, Universidad de Alicante, Campus de San Vicente del Raspeig, E-03690 Alicante, Spain.\\}

\author{A. Di Teodoro}
\email{nditeodoro@usfq.edu.ec}
\affiliation{Departamento de Matem\'aticas, Colegio de Ciencias e Ingenier\'ia, Universidad San Francisco de Quito,  Quito 170901, Ecuador\\}

\author{M. Mena}
\email{fmenal@estud.usfq.edu.ec}
\affiliation{Departamento de F\'isica, Colegio de Ciencias e Ingenier\'ia, Universidad San Francisco de Quito,  Quito , Ecuador\\}

\begin{abstract}
In this work, we introduce a new fractional derivative that modifies the conventional Riemann-Liouville operator to obtain a set of fractional Einstein field equations within a 2+1 dimensional spacetime by assuming a static and circularly symmetric metric. The main reason for introducing this new derivative stems from addressing the divergence encountered during the construction of Christoffel symbols when using the Caputo operator and the appearance of unwanted terms when using the Riemann-Liouville derivative because of the well-known fact that its action on constants does not vanish, as expected. The key innovation of the new operator ensures that the derivative of a constant is zero. As a particular application, we explore whether the Ba\~nados-Teitelboim-Zanelli black hole metric is a solution to fractional Einstein equations. Our results reveal that for values of the fractional parameter close to one, the effective matter sector corresponds to a charged BTZ solution with an anisotropic cosmological constant.
\end{abstract}
\maketitle
\section{Introduction}
One of the current challenges in physics is the reconciliation of gravity and quantum mechanics. Approaches such as the string theory \cite{1,2,3,4} and loop quantum gravity \cite{5,6} have been developed for a long time and appear to be the primary candidates for successful models in this endeavor but, it is undeniable that they also face conceptual and technical challenges, prompting the search for alternative formulations. Although the usual approach to quantizing gauge theories fails in the case of gravity owing to its lack of renormalizability, new alternatives that enable successful perturbative development through the use of non-local operators have recently been proposed. This is where fractional quantum gravity models, wherein the classical derivative is replaced by a fractional derivative, have emerged as a promising option \cite{Calcagni2012,7,8,9,10}. {In \cite{Calcagni2012}, the concept of a fractional spacetime is introduced, where the metric exists at all scales, but the Hausdorff and spectral dimensions vary with scale. Fractional calculus provides a natural framework for defining derivatives, volumes, and symmetries in these spaces. This implies that the spacetime metric is not a rigid structure but adapts to the observed scale, reflecting fractal properties in certain regimes.
In \cite{7}, the impact of multifractional measures on the notions of distance and volume in spacetime is analyzed. Despite these modifications, the metric remains present, although its operational meaning changes with scale. This work reinforces the idea that spacetime does not disappear at microscopic scales; instead, its geometric structure becomes multifractional, altering the metric relationships between events.
In \cite{9}, the author proposes a generalization of the Einstein-Hilbert action using fractional derivatives. This leads to modified field equations that, while still depending on an underlying metric, exhibit scale-dependent symmetries. This result demonstrates that the metric remains the fundamental object for defining gravity, but the equations governing it are modified by the use of fractional operators. The reasons are several, and here we list some of them.
In \cite{10}, the application of these ideas to early universe cosmology is explored. It is found that primordial fluctuations carry information about the multifractional structure of spacetime, enabling a connection between metric modifications and astrophysical observations. This result is crucial as it suggests that the scale-dependent metric structure is not merely a mathematical construction but has observable consequences in the cosmic microwave background and the quantum fluctuations of the early universe.} {In summary, fractional calculus naturally describes non-local phenomena, where a system's state at a point depends on its behavior over a region. In general relativity, spacetime curvature could be influenced by the surrounding energy-momentum distribution in a way that parallels non-local phenomena. Introducing fractional derivatives into Einstein's equations may modify their standard form, potentially adding terms that capture memory effects, hereditary behavior, and non-local deviations. Furthermore, fractional calculus might allow for dimensional interpolation between classical 3+1 spacetime and scale-dependent or fractional dimensions, possibly offering insights into quantum and emergent gravity. This approach could lead to a redefinition of curvature tensors, geodesics, and spacetime intervals within a fractional framework.}
In this regard, it would be pertinent to ask whether the non-locality introduced by fractional operators is merely a tool for constructing renormalizable theories or if it also makes sense at a classical level, that is, in the context of Einstein's equations, where it is known that problems such as the presence of essential singularities in black hole solutions and the existence of dark matter and dark energy remain and the non-locality could play an important role in alternative solutions for these issues.  However, this question poses significant conceptual challenges, as it completely loses the notion of spacetime from General Relativity and is only recovered in the limit where the fractional derivative coincides with the usual derivative. Nevertheless, models in which gravity is considered an emergent phenomenon are not new, and this could be the case, namely, that spacetime can be thought of as an entity that we only recover in certain regimes (of course, this is a conjecture as many others in the literature). However what is this regime and how do we connect the results to standard General Relativity? As we will see later, fractional derivatives depend on a fractional parameter. When this parameter equals zero, the fractional derivative is the identity (is the function again), and when it equals one, it yields the standard first derivative of the function. If we consider the fractional parameter to be close to one, we can think that the solutions of Einstein's equations are (approximately) those obtained in the standard case but with corrections given by the non-locality introduced by the fractional derivative. In other word, when we are ``close'' to the ordinary derivative, we can say that the geometry is the same, but the corrections are given by non-locality. This is analogous to what occurs when considering linearized gravity: the spacetime is Minkowski (indices are raised and lowered using the flat metric), but the perturbation of the gravitational field survives. Following the previous analysis, we can justify previous works aimed at extracting certain physics from the fractional version of Einstein's equations \cite{13,14,15,16,17,18,19} where fractional operators are used to some extent although none of them report exact solutions. In contrast, there are other works that start from effective actions involving a fractional parameter and have found modified Friedman equations, showing how the fractional parameter influences when comparing the model with observational data \cite{Garcia-Aspeitia:2022uxz,Gonzalez:2023who} {although the influence that fractional calculus is having on the mechanics of the universe was original introduced in \cite{Jalalzadeh:2020bqu,Moniz:2020emn,Jalalzadeh:2022uhl}. In \cite{Jalalzadeh:2020bqu}, the authors explore unconventional approaches to Quantum Cosmology (QC), focusing on the role of fractional calculus in modifying fundamental quantum gravity equations. A central contribution is the introduction of Riesz fractional derivatives for spatial differentiation in the Wheeler-DeWitt equation, replacing the standard Laplacian to incorporate nonlocal effects and fractal-like features of quantum spacetime. The Riesz derivative is chosen due to its preservation of isotropy, compatibility with quantum mechanics, and natural emergence in fractional quantum field theory. This formulation leads to a fractional Wheeler-DeWitt equation, where the fractional Laplacian modifies the structure of superspace and introduces scale-dependent corrections to the universe’s wave function evolution. The book extends this framework to black hole quantization, deriving a modified mass spectrum that incorporates fractional effects and suggesting a possible connection between fractional calculus and quantum gravitational thermodynamics. Overall, the work presents fractional derivatives as a fundamental extension of geometrodynamics, arguing that quantum gravity is inherently nonlocal and that observational imprints of fractional quantum cosmology (FQC) may appear in primordial fluctuations and the Cosmic Microwave Background (CMB).
In a related effort, \cite{Moniz:2020emn} explores the extension of FQM to QC by introducing fractional derivatives into fundamental equations governing the universe's evolution. The authors argue that standard quantum mechanics, based on the Brownian-Feynman path integral, can be generalized to include Lévy paths, leading to fractional Schrödinger equations with nonlocal effects. This approach is motivated by the fractal-like structure of quantum spacetime at Planckian scales, where the standard Laplacian is replaced by a Riesz fractional derivative to capture nonlocality while preserving rotational symmetry. In the context of QC, the authors propose a fractional Wheeler-DeWitt equation, modifying the minisuperspace dynamics by incorporating fractional spatial derivatives. The study explores the implications of this framework by applying it to a fractional FLRW cosmology, investigating how fractional quantum effects influence the flatness problem. It is shown that the presence of fractional corrections may eliminate the need for inflation by dynamically adjusting the universe’s initial conditions. The study also examines the role of fractional quantum mechanics in tunneling effects, arguing that a fractional approach modifies the probability amplitudes for universe nucleation. Overall, the paper establishes FQC as a novel framework, suggesting that nonlocal quantum effects, encoded through Riesz derivatives, could provide new insights into the initial conditions of the universe and the quantum nature of gravity. The authors emphasize that this approach remains in its early stages, requiring further mathematical and physical refinement, but it offers a promising direction for addressing fundamental problems in QC.
Further advancing this program, in \cite{Jalalzadeh:2022uhl}, the authors introduce a fractional Wheeler-DeWitt equation for a closed de Sitter universe, incorporating Riesz fractional derivatives to modify the standard quantum cosmological framework. By extending both the Linde-Vilenkin tunneling wavefunction and the Hartle-Hawking no-boundary proposal within this fractional setting, the authors demonstrate that the event horizon of the nucleated universe acquires a fractal structure, with dimensions constrained between $2\le D <3$. The tunneling wavefunction favors lower fractal dimensions ($D<2.5$), leading to an accelerated power-law expansion, whereas the no-boundary wavefunction corresponds to higher fractal dimensions $(D\approx3)$, resulting in a decelerated expansion. These results suggest that, due to the exponential suppression of closed universes in the tunneling approach, flat or open universes are preferred within a fractional inflationary framework. The authors justify the use of the Riesz derivative for spatial differentiation due to its natural extension of the Laplacian in fractional quantum mechanics, preserving isotropy and nonlocality, while ensuring a well-defined fractional extension of quantum cosmology. Furthermore, the study introduces a fractional entropy relation, linking the fractal horizon area with a modified Gibbons-Hawking entropy, implying that fractional quantum effects can alter the thermodynamics of de Sitter space. These results provide a novel perspective on the emergence of inflation and quantum spacetime, emphasizing the role of nonlocality and fractality in the early universe.}

Difficulties in applying fractional calculus to exactly solve Einstein's equations arise from the outset. The first question is, what are the fractional Einstein equations? In Ref. \cite{9}, the first step in constructing fractional Einstein equations is proposed, which involves substituting ordinary derivatives with fractional derivatives when defining the Christoffel symbols, namely
\begin{eqnarray}\label{SC}
\Gamma^{\mu}_{\nu\lambda}=\frac{1}{2}g^{\mu\delta}(D_{\nu}g_{\delta\lambda}+
D_{\lambda}g_{\delta\nu}-
D_{\delta}g_{\nu\lambda}), 
\end{eqnarray}
where $\mu,\nu=0,1,2,3$ are space-time indices and $D_{\nu}$ stands for the fractional derivative with respect to the coordinate $\nu$ which in the classical limit coincides with the usual derivative operator, namely $D_{\nu}\to\partial_{\nu}$ (details about this limit can be found in the next section). Note that if we define the fractional covariant derivative $\tilde{\nabla}$
of a  $(k,l)$-rank tensor $T$ as

\begin{equation}\label{CD}
\tilde{\nabla}_\sigma T^{\mu_1\mu_2\cdots\mu_k}_{\nu_1\nu_2\cdots\nu_l} = D_\sigma T^{\mu_1\mu_2\cdots\mu_k}_{\nu_1\nu_2\cdots\nu_l} + \Gamma^{\mu_1}_{\sigma\lambda} T^{\lambda\mu_2\cdots\mu_k}_{\nu_1\nu_2\cdots\nu_l} + \Gamma^{\mu_2}_{\sigma\lambda} T^{\mu_1\lambda\cdots\mu_k}_{\nu_1\nu_2\cdots\nu_l} + \cdots - \Gamma^\lambda_{\sigma\nu_1} T^{\mu_1\mu_2\cdots\mu_k}_{\lambda\nu_2\cdots\nu_l} - \Gamma^\lambda_{\sigma\nu_2} T^{\mu_1\mu_2\cdots\mu_k}_{\nu_1\lambda\cdots\nu_l} - \cdots
\end{equation}
it is compatible with the metric in the sense that \cite{9}
\begin{eqnarray}\label{compatible}
\tilde{\nabla}_{\mu}g_{\nu\rho}=0.    
\end{eqnarray}

The Riemann tensor, Ricci tensor, and Ricci scalar can now be defined as \cite{9}
\begin{eqnarray}
R_{\rho \mu \sigma \nu} &=& D_{\sigma} \Gamma^{\alpha}_{\mu \nu} - D_{\nu} \Gamma^{\alpha}_{\mu \sigma} + \Gamma^{\tau}_{\mu \nu} \Gamma^{\alpha}_{\sigma \tau} - \Gamma^{\tau}_{\mu \sigma} \Gamma^{\alpha}_{\nu \tau}\label{riemann}\\
R_{\mu\nu}&=& D_{\sigma} \Gamma^{\sigma}_{\mu\nu}
-D_{\nu} \Gamma^{\sigma}_{\mu\sigma}
+\Gamma^{\tau}_{\mu\nu}\Gamma^{\sigma}_{\sigma\tau}-
\Gamma^{\tau}_{\mu\sigma}\Gamma^{\sigma}_{\nu\tau}\label{Ricci0}\\
R&=&g^{\mu\nu}R_{\mu\nu}\label{scalar0}.
\end{eqnarray}
It is worth emphasizing that Eqs. (\ref{SC})-(\ref{scalar0}) correspond to straightforward generalizations because we only replaced the usual derivative operator with the fractional one. Indeed, these expressions serve as a good starting point because, as mentioned earlier, we are assuming that the ``standard'' geometry only emerges in the limit where $D_{\mu}$ coincides with $\partial_{\mu}$. Of course, it would be interesting to start from first principles, such as considering the definition of the Riemann tensor
\begin{equation}
R(X,Y) = [\tilde{\nabla}_{X},\tilde{\nabla}_{Y}] - \nabla_{[X,Y]}
\end{equation}
where $[X, Y]$ is the Lie bracket of vector fields and $[\nabla_{X},\nabla_{Y}]$ is the commutator of differential operators. However, this raises other questions about what fractional Lie brackets are or how to generalize the commutator between vector fields in such a way that yields (\ref{riemann}). Another question is whether these commutators obey a certain fractional Jacobi identity that leads to the Bianchi identity, namely
\begin{eqnarray}\label{formalbianchi}
\tilde{\nabla}_{[\alpha}R^{\lambda}_{\beta\gamma]\delta}=0.
\end{eqnarray}
Answering these questions is not trivial, given the lack of the usual Leibniz rule for fractional operators.  Indeed, a Leibniz rule exists for fractional derivatives, but it contains an infinite series of classical derivatives of the functions involved, making it difficult to handle \cite{KST2006, P1999, SKM1993}. The intention of this work is not to answer these questions but rather to leave open the possibility of exploring these details in future work.\\

Now, by considering (\ref{SC})-(\ref{scalar0}) as the starting point, we have two routes to obtain Einstein's field equations: i) solving the variational problem from the Einstein-Hilbert action or ii) considering the Einstein field equations as a starting point by using (\ref{Ricci0}) and (\ref{scalar0}). If we start from the Einstein-Hilbert action, we encounter, after variations with respect to the metric, the boundary term
\begin{eqnarray}
\delta g^{\mu\nu}\mathcal{O}_{\mu\nu}=\nabla_{\sigma}(g^{\mu\nu}\delta\Gamma^{\sigma}_{\mu\nu}-g^{\mu\sigma}\delta\Gamma^{\rho}_{\mu\rho})
\end{eqnarray}
that vanishes in the case of ordinary derivatives but generally does not vanish when fractional operators are used (see \cite{9} for details). By contrast, if we consider the Einstein field equations as fundamentals, the boundary term is (apparently) absent, and we have
\begin{eqnarray}\label{EFE}
G_{\mu\nu}=R_{\mu\nu}-\frac{1}{2}R =\kappa^{2}T_{\mu\nu}.
\end{eqnarray}
Note that if (\ref{formalbianchi}) is true, is straightforward to show that $\tilde{\nabla}_{\mu}(R^{\mu\nu}-\frac{1}{2}g^{\mu\nu}R)=0$, from which $\tilde{\nabla}_{\mu}T^{\mu\nu}=0$ which means that the energy-momentum tensor in (\ref{EFE}) is conserved in the fractional sense. Regardless of the case, in this work, we will consider (\ref{EFE}) as true meaning that, in order to be compatible with the equations obtained from the Einstein-Hilbert action, the energy-momentum tensor must be conceived as a quantity that contains the matter sector and the corrections introduced by $\mathcal{O}_{\mu\nu}$, that is,
\begin{eqnarray}
T_{\mu\nu}\to\ T_{\mu\nu}+\mathcal{O}_{\mu\nu}.
\end{eqnarray}
We must emphasize that, strictly speaking, in order to have vacuum solutions, it is necessary to set $T_{\mu\nu} = 0$ consistently, which would lead to a constraint on the metric given by ${\cal O}_{\mu\nu} = 0$, or to a constraint on the matter distribution $T_{\mu\nu} = -{\cal O}_{\mu\nu}$, given a certain metric. However, in this work, we are not interested in finding vacuum solutions, but rather in providing a metric to obtain the effective matter sector (which contains information about both $T_{\mu\nu}$ and $\mathcal{O}_{\mu\nu}$).
\\

Now, the problem becomes finding solutions to the field equations, which in this case translates to dealing with a system of 10 integro-differential equations as each fractional derivative operator involves an integral, as shown in the next section. The standard approach to reducing the number of equations to be solved is to assume a system with certain symmetries and a particular parameterization for the metric. For instance, in $3+1$ spacetime dimensions, the simplest assumption is a spherically symmetric and static system, whose line element is given by
\begin{equation}\label{metricfull}
ds^{2}=-F(r)dt^{2}+G(r)dr^{2}+r^{2}d\theta^{2}+r^{2}\sin^{2}\theta d\phi^{2}.
\end{equation}
As is known (and expected), Einstein equations depend only on the radial coordinate. However, it is worth mentioning that some non-zero Christoffel symbols explicitly contain the coordinate $\theta$, which does not appear in the final result owing to Leibniz's rule and trigonometric identities. However, as stated above, in fractional calculus there is no standard Leibniz's rule but a relation involving an infinite series in derivatives which is not trivial to deal with. In this regard, we have to either find the convergence of the series (which is not trivial) or impose constraints to ensure the classical limits. For example, in the case of (\ref{metricfull}) we encounter the terms
\begin{eqnarray}
D_{3}\left(\frac{D_{3}\sin^{2}\theta}{\sin^2{2}\theta}\right)+\frac{D_{3}(D_{3}\sin^{2}\theta)}{\sin^{2}\theta}&=&-2\\
\left(\frac{D_{3}\sin^{2}\theta}{\sin^{2}\theta}\right)^{2}-D_{3}\left(\frac{D_{3}\sin^{2}\theta}{\sin^2{2}\theta}\right)+\frac{D_{3}(D_{3}\sin^{2}\theta)}{\sin^{2}\theta}&=&0,
\end{eqnarray}
Where $D_{3}$ is the fractional derivative with respect to $\theta$, which comes from the computation of $G_{22}$ and $G_{33}$ of the Einstein tensor (note that, if we assume $D_{3}=\partial_{3}=\partial/\partial\theta$, the constraints are automatically satisfied). One way to surpass this ``polar angle problem'' is by studying a toy model such as Einstein's equations in $2+1$ dimensional spacetime, where for circularly symmetric and static situations, the metric reads
\begin{eqnarray}\label{metric3}
 ds^{2}=-F(r)dt^{2}+G(r)dr^{2}+r^{2}d\phi^{2}
\end{eqnarray}
{It should be emphasized that such a dimensional reduction is not related to any fundamental aspect, but the problem is more manageable from a technical point of view.} However, in this case, we have to face another problem (which is also present if we use (\ref{metricfull})) involving the operation $D_{1}(r^{-2}D_{1}r^{2})$ with $D_{1}$ as the fractional derivative with respect to the radial coordinate, which appears as a consequence of the component $g_{22}=r^{2}$ in (\ref{metricfull}) when constructing the Ricci tensor. For example, it can be seen that this term does not converge for all values of $r$ if we implement the widely used Caputo derivative \cite{KST2006, P1999, SKM1993}. This issue could be addressed if we utilize the Riemann-Liouville derivative \cite{KST2006, P1999, SKM1993}, for instance, but we encounter the problem that, since its action on constants does not vanish as expected, it leads to the appearance of unwanted terms in the Einstein equations. More specifically, the Einstein equations for a static and circularly symmetric metric in 2+1 dimensions could ultimately depend on the time and axial angle.\\

In this work, we focus on addressing these issues by defining a new derivative, which is a modification of the Riemann-Liouville fractional derivative and possesses the property that the derivative of a constant is zero. As we will see later, this new derivative contains weights that are conveniently introduced to regularize problematic points and to obtain the set of Einstein equations for a given parameterization of the metric. In our case, this corresponds to three integro-differential equations with five unknowns. It should be noted that the techniques to solve this system of coupled integro-differential equations are beyond the scope of this work. Instead, we propose a form of the metric and evaluate how the fractional operators affect the material sector. More precisely, we propose that the metric is the Ba\~nados-Teitelboim-Zanelli (BTZ) vacuum solution of the non-fractional (standard) equations, and we find that for BTZ to be a solution of the fractional equations, the material sector cannot be empty but rather a kind of non-uniform cosmological constant. We want to emphasize that this work is a first approach where we highlight the problems that the usual operators have in obtaining the fractional Einstein equations and how the modified weights might help solve these problems. At no point do we claim that our derivative is the most optimal; instead, we leave the door open for this idea to be explored by the interested community.\\

This work is organized as follows. The next section, introduces the formal aspects of the work. Next, in Section \ref{EFEs}, we deduce the set of Einstein field equations for a static and circularly symmetric $2+1$ dimensional spacetime. In particular, we explore the consequences of assuming the Ba\~nados-Teitelboim-Zanelli (BTZ) metric \cite{BTZ1,BTZ2} as a solution to the fractional equations and explore its consequences. Finally, the last section is devoted to some final comments and remarks.

\section{Formalism}
In this section, we present the basic definition and notation related to fractional calculus used in this study. As we aim for this work to be as self-contained as possible, we will start with the definitions of the Caputo and Riemann-Liouville derivatives to lay the groundwork for our new fractional derivative.\\

Let us start with the definition of the Riemann integral which plays a central role in the definition of the fractional derivatives  used here.

\begin{defn}
The Riemann--Liouville fractional integral of order $\eta>0$ is given by (see \cite{KST2006, P1999, SKM1993})
\beqn\label{IRLO}
\left(I_{a^+}^\eta h \right)(x) ~=\frac{1}{\Gamma(\eta)} \int_a^x \frac{h(t)}{(x-t)^{1-\eta}} ~dt, \quad x>a. \label{pre3}
\eeqn

\noindent We denote by $I_{a^+}^\eta(L_1)$ the class of functions $h$, represented by the fractional integral (\ref{IRLO}) of a summable function, that is $h=I_{a^+}^{\eta}\varphi$, where $\varphi \in L_1(a,b).$ A description of this class of functions was provided in \cite{KST2006, SKM1993, stein}.
\end{defn}

\begin{thm}\label{Thm1}
A function $h \in I_{a^+}^\eta(L_1), \eta>0$, if and only if its fractional integral $I_{a^+}^{s-\eta} h \in AC^s([a,b])$, where $s=[\eta]+1$ and
$(I_{a^+}^{s-\eta} h )^{(k)}(a)=0$, for $k=0,\ldots,s-1$. 
\label{Th1_Pre}
\end{thm}

\noindent In Theorem \ref{Thm1}, $AC^s([a,b])$ denotes the class of functions $h$, which are continuously differentiable on the segment $[a,b]$, up to order $s-1$ and $h^{(s-1)}$ is absolutely continuous on $[a,b]$. By removing the last condition in Theorem \ref{Thm1}, we obtain a class of functions that admit a summable fractional derivative. (See \cite{KST2006, SKM1993})
\begin{defn}[ see \cite{SKM1993}]
A function $h \in L_1(a,b)$ has a summable fractional derivative $\left(D_{a^+}^\eta h \right)(x)$ if

$\left(I_{a^+}^{s-\eta} h \right)(x) \in AC^s([a,b])$,

where $s=[\eta]+1.$
\end{defn}


\begin{defn}
Let $\left(D_{a^+}^\eta h \right)(x)$ denote the {\bf fractional Riemann--Liouville derivative} of order $\eta>0$ (see \cite{KST2006, P1999,  SKM1993})
\begin{align}\label{fracderivative}
\left(D_{a^+}^\eta h \right)(x) &=\left(\frac{d}{dx}\right)^s \frac{1}{\Gamma(s-\eta)}\int_a^x \frac{h(t)}{(x-t)^{\eta-s+1}} ~dt\nonumber\\
&=\left(\frac{d}{dx}\right)^s \left(I_{a^+}^{s-\eta} h \right)(x),
\end{align}
where $s=[\eta]+1, x>a$ $[\eta]$ denotes the integer part of $\eta$ and $\Gamma$ is the gamma function. When $0 <\eta <1$ , then (\ref{fracderivative}) takes the form
\end{defn}
\beqn\label{RLO}
\left(D_{a^+}^\eta h \right)(x)~ =\frac{d}{dx} \left(I_{a^+}^{1-\eta} h \right)(x).
\eeqn
Note that, when $\eta\to1$, we recover the typical derivative operator \cite{KST2006, P1999,  SKM1993}.
\\

\noindent The semigroup property for the composition of fractional derivatives does not generally hold (see \cite[Sect. 2.3.6]{P1999}). In fact, the property:
\beqn
D_{a^+}^\eta\left(D_{a^+}^\gamma h \right) =D_{a^+}^{\eta+\gamma} h \label{EL},
\eeqn
holds whenever
\beqn
h^{(j)}(a^+) =0, \qquad j=0,1, \ldots, s-1, \label{ELC}
\eeqn
and $h \in AC^{s-1}([a,b])$, $h^{(s)} \in L_1(a,b)$ and $s=[\gamma]+1$. Thus, we can write this result in the following:

\begin{lem}\label{lemvital}
Consider $h\in AC^{s-1}([a,b])$ and $h^{(s)}\in L_{1}(a,b)$ then, 

\begin{equation}
\dsp D_{a+}^{\eta}\left(  \dsp D_{a+}^{\gamma}\right) h =\dsp D_{a+}^{\gamma}\left( \dsp D_{a+}^{\eta}\right) h,
\end{equation} 
holds whenever

\begin{equation}
h^{(j)}(a^{+})=0, \quad j=0,1,\dots, s-1,
\end{equation}
where $s=[\gamma]+1$.

\begin{proof}
This proof can be found in \cite[Secton 2.3.6]{P1999}.
\end{proof}
\end{lem}

\begin{rem}
It is worth noticing that the {\bf Riemann-Liouville derivative of a constant is not zero}. However, in the limit process, it behaves as expected.
\begin{equation}
\dsp \lim_{\eta\to 1}\left(D_{a^+}^\eta 1 \right)(x)=\lim_{\eta\to 1}\frac{(x-a)^{-\eta}}{\Gamma(1-\eta)}=0.
\end{equation}
\end{rem}

\begin{defn}\label{deffc} Let $\eta \geq 0$ and $m = [\eta]$. We can then define {\bf the Caputo derivative} \cite{KST2006, P1999,  SKM1993} 
$_{c}{D}_{a^{+}}^{\eta}$ as 
\begin{equation}
_{c}{D}_{a^{+}}^{\eta}f = I_{a^{+}}^{m-\eta}\left( \dsp \frac{d}{dx} \right)^{m}f, 
\end{equation}
when $\left( \dsp \frac{d}{dx} \right) ^{m}f \in L_{1}[a,b]$. 
\end{defn}

Note that Caputo derivative of a constant is zero, as expected. This presents an advantage when constructing Christoffel symbols because, for example, if the metric depends only on the radial coordinate, its derivative with respect to time vanishes (something that would not occur when using the Riemann-Liouville derivative).  However, the use of Caputo is problematic when constructing the term
\begin{eqnarray}\label{problematic0}
{}_{c}D^{\gamma}_{0^{+}}(r^{-2}{}_{c}D^{\eta}_{0+}r^{2}) =-\frac{2\eta}{\Gamma(1-\eta) \Gamma(3-\eta)} \int_0^r \frac{t^{-\eta-1}}{(r-t)^{\gamma}}dt    
\end{eqnarray}
because it fails to converge when $r\to0$ and $\eta>0$. One option is to contemplate the Riemann integral from a certain point $a>0$. However, this approach introduces additional challenges because the solution becomes approximate rather than exact, necessitating numerical computation for a given fixed $a$. Moreover, when employing standard derivatives, this issue is absent, and our aim here is to ensure that our equations inherit all the desirable properties. 
Now, although the Riemann-Liouville derivative does not diverge in this case, it is not useful for deriving the Einstein equations, because, as we saw earlier, the derivative of a constant is not zero. For this reason, we propose a modification to the Riemann-Liouville operator such that its action on a constant vanishes, as we see in what follows.

\begin{defn}[]
Consider $q_{1}(x,\eta)$ a continuous function, $q_{2}(x,\eta)$ a continuously differentiable function on $x$ and let $\left({}^{(q_{1},q_{2})}D_{a^+}^\eta h \right)(x)=\left({}^{\overline{q}}D_{a^+}^\eta h \right)(x)$ denote the {\bf $\overline{q}$-weighted fractional Riemann-Liouville derivative}  of order $\eta>0$. For $q_{1},q_{2}\in AC^{s}(\mathbb{R})$
\begin{equation}
\left({}^{\overline{q}}D_{a^+}^\eta h \right)(x)= 
q_{1}(x,\eta)\left(\frac{d}{dx}\right)^s q_{2}(x,\eta)\left(I_{a^+}^{s-\eta} h \right)(x),
 \label{pre2}
\end{equation}
where $s=[\eta]+1, x>a$ and $[\eta]$ denotes the integer part of $\eta$
with $0 <\eta <1$ and $\dsp \lim_{\eta\to 1} q_{1}(x,\eta) =\lim_{\eta\to 1} q_{2}(x,\eta)=1$. As we will see later, for convenience we take $q_{2}(x,\eta)=(x-a)^{\eta-1}$ so then (\ref{pre2})  takes the form
\begin{align}
\left({}^{\overline{q}}D_{a^+}^\eta h \right)(x)= 
q_{1}(x,\eta)\left(\frac{d}{dx}\right) (x-a)^{\eta-1}\left(I_{a^+}^{1-\eta} h \right)(x). 
\end{align}
\end{defn}
It is not diffficult to see that the operator is linear; however, the central motivation for considering this operator is that, although it has a structure similar to that of the Riemann-Liouville operator, the derivative of a constant is zero
\begin{align}
\left({}^{\overline{q}}D_{a^+}^\eta 1 \right)(x)=0
\end{align}
Another consequence arises when we differentiate polynomials. If we consider $a=0$, $P(x)=x^n$ and $x,n>0$, 
\begin{align}
\left({}^{\overline{q}}D_{0^+}^\eta P(x) \right)(x)=\frac{n q_{1}(x,\eta) x^{n-1} \Gamma (n+1)}{\Gamma (n-\eta +2)},
\end{align}
and $\dsp \lim_{\eta\to 1}\left({}^{\overline{q}}D_{0^+}^\eta P(x) \right)(x)= n  x^{n-1}$.

As in the case of the Riemann-Liouville operator, the semigroup property for the composition of fractional derivatives is generally not satisfied; however, the following lemma is useful

\begin{lem}\label{lemvital2}
Consider $q_{2}(x,\eta)\in AC^{s}([a,b]\times (0,1))$, $h\in AC^{s-1}([a,b])$ and $h^{(s)}\in L_{1}(a,b)$ then, 

\begin{equation}
\dsp {}^{\overline{q}}D_{a^+}^{\eta}\left(  \dsp {}^{\overline{q}}D_{a^+}^{\gamma}\right) h =\dsp {}^{\overline{q}}D_{a^+}^{\gamma}\left( \dsp {}^{\overline{q}}D_{a^+}^{\eta}\right) h,
\end{equation} 
holds whenever

\begin{equation}
h^{(j)}(a^{+})=0, \quad j=0,1,\dots, s-1,
\end{equation}
where $s=[\gamma]+1$.

\begin{proof}
The proof can be followed using the formula 

\begin{align}
\dsp \left({}^{\overline{q}}D_{a^+}^\eta h \right)(x)=
q_{1}(x,\eta)\Big[\left(\frac{d}{dx}\right)^s q_{2}(x,\eta)\Big]\left(I_{a^+}^{s-\eta} h \right)(x)
+
q_{2}(x,\eta)\left(D_{a^+}^\eta h \right)(x)
\end{align}
and the result of lemma \ref{lemvital}
\end{proof}
\end{lem}

Note that, up to this point, the weight $q_{2}$ is general. However, to avoid singularities in (\ref{problematic0}), it is convenient to consider
\begin{eqnarray}\label{pesos}
q_{2}(x,\eta)=\frac{1}{q_{1}(x,\eta)}=(x-a)^{1-\eta}.    
\end{eqnarray}
Indeed, in this case we obtain
\begin{eqnarray}\label{problematic0.1}
\dsp {}^{\overline{q}}D^{\gamma}_{0^{+}}(r^{-2}\ \dsp {}^{\overline{q}}D^{\eta}_{0+}r^{2})=
-\frac{4r^{-\gamma-\eta}\gamma\Gamma(1-\eta)}{\Gamma(4-\eta)\Gamma(2-\eta-\gamma)}
\end{eqnarray}
At this point, some comments are in order. First, note that we have used different fractional indices that translates to using $\eta$ to define the Christoffel symbols in (\ref{SC}) and $\gamma$ to define the curvature tensor (\ref{riemann}). Indeed, this is the most general way to consider fractional Einstein equations. Furthermore, if they are considered equal a priori, the result of (\ref{problematic0.1}) does not lead to the result obtained with the ordinary derivative in the limit. Second, the result obtained with the classical derivative is achieved if $\gamma\to1$ is taken before $\eta\to1$, otherwise, the expression diverges. Of course, this introduces a hierarchy in how the classical limit is recovered: first turning off non-locality in curvature, and then in the Christoffel symbols. Third, our definition of the weighted derivative somewhat resembles the weighted derivatives introduced in gravitational theories (see \cite{7,8} for a review). However, these derivatives acquire an undesired property to some extent, in that the weighted derivative of a constant is not zero as one would expect. In the context of fractional calculus, the Riemann-Liouville derivative of a constant is not zero, but we have introduced the weight to make the derivative of a constant now be zero. {Finally, our operator satisfies a generalized non-trivial Leibniz rule from the Riemann-Liouville derivative. For the standard Riemann-Liouville fractional derivative, the Leibniz rule for the product of two functions $f(x)$ and $g(x)$ is given by:
\begin{equation}
D^\eta_{a^+} [f(x) g(x)] = \sum_{k=0}^{\infty} \binom{\eta}{k} (D^k_{a^+} f(x)) (D^{\eta-k}_{a^+} g(x)).
\end{equation}
where $\binom{\eta}{k}$ is the generalized binomial coefficient.
In our case, the modified fractional derivative is defined as:
%
\begin{equation}
{}^{\overline{q}}D^\eta_{a^+} [f(x) g(x)] = q_{1}\left(\frac{d}{dx} q_2(x,\eta)\right)I_{a^+}^{1-\eta-k} (f(x) g(x))+q_{1}q_{2}\sum_{k=0}^{\infty} \binom{\eta}{k} (D^k_{a^+} f(x)) (D^{\eta-k}_{a^+} g(x)).
\end{equation}
}
\\

In the next section, we derive the Einstein field equation for a static and circularly symmetric metric by assuming the weights in (\ref{pesos}) and taking $a=0$. To make the equations readable, we use the notation
\begin{eqnarray}\label{notation}
({}^{\overline{q}}D_{0+}^{\eta})_{r}\equiv d^{\eta}.
\end{eqnarray}

{Before concluding this section, it is worth discussing the value of the fractional parameter \(\eta\). In this work, we restrict the parameter to the range \(0 < \eta < 1\). While it is well-established that corrections arising from quantum gravity typically require \(\eta > 1\), the corrections considered here may originate from classical non-locality. This non-locality manifests in the theory through mechanisms that are not yet fully understood but remain an active area of investigation.
We hypothesize that there could be a mechanism analogous to the one proposed in \cite{Narasimhan1993}, where the significance of long-range or non-local effects in the mechanical properties of materials is attributed to stress being a function of the mean strain over a representative volume of the material centered at a given point. Similarly, in our context, the fractional parameter \(\eta\) could encode non-local interactions with a classical origin, offering an intriguing avenue for further exploration.
}
\\

{Let us summarize our finding in this section. We introduce the necessary definitions and notation for the fractional formulation of Einstein's equations. We discuss the most commonly used fractional derivatives, namely the Riemann-Liouville and Caputo derivatives, and the challenges they pose when applied to general relativity. 
The Riemann-Liouville derivative has the drawback that it does not vanish when acting on a constant function, which may introduce spurious degrees of freedom in the field equations. On the other hand, the Caputo derivative, while ensuring that the derivative of a constant is zero, leads to divergences in the construction of the Christoffel symbols when applied in the domain of the radial coordinate $r$. This failure makes it unsuitable for our framework.
To overcome these issues, we introduce a modified version of the Riemann-Liouville derivative. This new fractional derivative possesses two crucial properties: it vanishes on constants, preventing spurious degrees of freedom, and it avoids the divergences that arise when constructing the Christoffel symbols using the Caputo derivative. 
To achieve this, we introduce weighting functions $q_1(x, \eta)$ and $q_2(x, \eta)$ in the definition of the fractional operator. In particular, we choose a weight that depends on the radial coordinate, ensuring a consistent and divergence-free formulation. 
This approach allows us to derive a fractional formulation of Einstein's equations that preserves the fundamental structure of the theory while incorporating nonlocal effects in a controlled manner. In the next section, we apply this framework to explicitly derive the fractional Einstein equations in a 2+1 dimensional static and circularly symmetric spacetime.
}

\section{Fractional Einstein's equations in 2+1 dimensions}\label{EFEs}
In what follows, we obtain the explicit form of the Einstein field equations for the $2+1$ dimensional metric
\begin{eqnarray}\label{metric0}
ds^{2}=-F(r)dt^2 +G(r)dr^2 + R(r)d\phi^2,   
\end{eqnarray}
from (\ref{SC}), (\ref{Ricci0}), (\ref{scalar0}), and (\ref{EFE}), with the convention that the index $\eta$ is used in the definition of the Christoffel symbols and $\gamma$ in the definition of the Riemann tensor. Note that, because the metric is a function of the radial coordinate, only terms involving the fractional derivative with respect to the radius will persist.  Using (\ref{notation}), the fractional Einstein field equations can be written as

\begin{eqnarray}
   -T^{0}_{0} &=& -\frac{(d^{\eta }F)^2}{32 \pi  F^2 G}+\frac{d^{\eta }F d^{\eta }G}{32 \pi  F G^2}+\frac{d^{\gamma}\left(G^{-1}d^{\eta }F\right)}{32 \pi  F}-\frac{d^{\gamma }\left(F^{-1}d^{\eta }F\right)}{32 \pi  G}\nonumber\\
   &&-\frac{d^{\gamma }\left(R^{-1}d^{\eta }R\right)}{32 \pi  G}-\frac{d^{\gamma}\left(G^{-1}d^{\eta }R\right)}{32 \pi  R} \label{EFE1}\\
T^{1}_{1}&=& -\frac{(d^{\eta }F)^2}{32 \pi  F^2 G}+\frac{d^{\eta }F d^{\eta }G}{32 \pi  F G^2}+\frac{d^{\gamma }\left(G^{-1}d^{\eta }F\right)}{32 \pi  F}-\frac{d^{\gamma }\left(F^{-1}d^{\eta }F\right)}{32 \pi  G}+\frac{d^{\eta }F d^{\eta }R}{32 \pi  F G R}\nonumber\\
&&+\frac{d^{\eta }G d^{\eta }R}{32 \pi  G^2 R}-\frac{(d^{\eta }R)^2}{32 \pi  G R^2}+\frac{d^{\gamma }\left(G^{-1}d^{\eta }R\right)}{32 \pi  R}-\frac{d^{\gamma }\left(R^{-1}d^{\eta }R\right)}{32 \pi  G}\label{EFE2}\\
T^{2}_{2}&=&\frac{d^{\gamma }\left(F^{-1}d^{\eta }F\right)}{32 \pi  G}+\frac{d^{\gamma }\left(G^{-1}d^{\eta }F\right)}{32 \pi  F}-\frac{d^{\eta}G d^{\eta }R}{32 \pi  G^2 R}+\frac{(d^{\eta }R)^2}{32 \pi  G R^2}\nonumber\\
&&+\frac{d^{\gamma}\left(R^{-1}d^{\eta }R\right)}{32 \pi  G}-\frac{d^{\gamma }\left(G^{-1}d^{\eta }R\right)}{32 \pi  R}.\label{EFE3}
\end{eqnarray}

It is worth noticing that the dimensions of the field equations change depending on the choice of weights  $(q_{1},q_{2})$. However, with the choice in equation (\ref{pesos}), the dimensions of the fractional derivative of the metric are length$^{-\eta}$. Consequently, the dimensions of the Einstein equations are length$^{-\eta-\gamma}$ so thus, when $\eta\to1$ and $\gamma\to1$ we obtain the correct dimensions (in natural units). From a physical standpoint, this means that we must be careful when assigning the meaning of each component of the fractional energy-momentum tensor $T_{\mu\nu}$. More precisely, in the standard case, we assign 
$T^{\mu}_{\nu}=(-\rho, p_{r},p_{t})$ with $\rho$, $p_{r}$ and $p_{t}$ as the energy density, radial pressure, and tangential pressure, respectively, and each quantity has dimensions of length$^{-2}$ but when applying fractional operators it is convenient to define
\begin{eqnarray}
T^{0}_{0}&=&-\Xi_{1}\rho\\ 
T^{1}_{1}&=&\Xi_{2}p_{r} \\
T^{2}_{2}&=&\Xi_{3}p_{t}
\end{eqnarray}
with $\{\Xi_{i}\}$, some quantity with dimensions of length$^{2-\eta-\gamma}$ (which can be taken as a constant associated with some characteristic length of the model under study) so that $\{\rho,p_{r},p_{t}$\} have the correct dimensions.\\

The next step in the program is to solve the set of fractional derivatives
(\ref{EFE1})-(\ref{EFE3}) with $R=r^2$  so we must solve the problem of solving three integro-differential equations with five unknowns, namely, 
$\{\rho,p_{r},p_{t},F,G\}$ which represents a non-trivial challenge. We can try by following the routes we explore in standard General Relativity, namely
\begin{enumerate}
    \item Provide an equation of state relating $\rho$ and $p_{r}$ and a suitable anisotropic function.
    \item Consider a matter sector based on fundamental fields.
    \item Provide some geometric restrictions
    \item Consider a vacuum solution.
\end{enumerate}

Regarding the last point, the only non-trivial vacuum solution for the standard $2+1$ Einstein field equations with a negative cosmological constant was found in \cite{BTZ1,BTZ2} whose metric, in the static and circularly symmetric regime, reads
\begin{eqnarray}\label{btzmetric}
F=\frac{1}{G}=-M+\frac{r^{2}}{\ell^{2}},    
\end{eqnarray}
which corresponds to the static  BTZ black hole with event horizon $r_{H}=\ell \sqrt{M}$. Alternatively, the BTZ black hole can be thought of as a solution of Einstein's equation without cosmological constant but supported by a matter sector given by $T^{\mu}_{\nu}=\frac{1}{8\pi \ell^{2}}diag(1,1,1)$. In this work, instead of trying to solve the set (\ref{EFE1})-(\ref{EFE3}), we assume (\ref{btzmetric})  as a solution to the fractional equations and
explore the behavior of the corresponding fractional matter sector. Before proceeding with the calculation, we would like to point out that the choice of a metric that solves the classical (non-fractional) Einstein equations is made for simplicity. One could attempt to use a metric that depends on the fractional parameters and coincides with BTZ in the appropriate limit. However, although we explored this approach, we did not find a metric simpler than the BTZ.\\ 

After using (\ref{btzmetric}) in the fractional equations, we obtain expressions for the matter sector that are not included here due to their length, as they involve hypergeometric functions and are not particularly illuminating. Furthermore, since our focus is on the behavior near the realm of general relativity, we consider the solution’s behavior for $\eta, \gamma$ close to one. By expanding in series for these parameters, we find that
\begin{eqnarray}
-T^0_0&=&\frac{1}{8 \pi  \ell^{2}}\left(1+\frac{\gamma-1}{2(\eta-1)}\right) +\frac{M(\gamma-1)}{16 \pi  r^2(\eta-1)}  + O\left(\eta-1, \gamma-1\right)\label{T00}\\ 
T^1_1&=&\frac{1}{8 \pi  \ell^{2}}\left(1-\frac{\gamma-1}{2(\eta-1)}\right) -\frac{M(\gamma-1)}{16 \pi  r^2(\eta-1)}  + O\left(\eta-1, \gamma-1\right)\label{T11}\\
T^2_2&=&\frac{1}{8 \pi  \ell^{2}}\left(1+\frac{\gamma-1}{2(\eta-1)}\right) +\frac{M(\gamma-1)}{16 \pi  r^2(\eta-1)}  + O\left(\eta-1, \gamma-1\right).\label{T22}
\end{eqnarray}

At this point some comments are in order. First, note that expressions (\ref{T00}-\ref{T22}) underscore the critical requirement of preserving hierarchy in the fractional parameters when taking the limit to recover the classic solution. Notably, it is essential that $\gamma$ approaches 1 prior to $\eta$, as $\gamma$ must reach this value first to ensure the correct limiting process. Second, around this limit, the introduction of non-locality leads to a small correction in the cosmological constant. In fact, the asymptotic behavior corresponds to a solution with an anisotropic cosmological constant. Finally, the second term in the matter sector $\propto r^{-2}$ is reminiscent of the charged BTZ black hole solution with line element \cite{MTZ}\\
\begin{eqnarray}\label{mtzmetric}
F=\frac{1}{G}=-M+\frac{r^{2}}{\ell^{2}}-\frac{Q^2}{2}\ln{\frac{r}{r_0}},    
\end{eqnarray}
 where $Q$ represents the electric charge of the black hole and $r_0$ is an arbitrary reference scale. Solving the Einstein Field Equations for this metric the matter sector is\\
\begin{eqnarray}
-T^0_0&=&\frac{1}{8 \pi  \ell^{2}} +\frac{Q^2}{32 \pi  r^2}  \label{T00q}\\ 
T^1_1&=&\frac{1}{8 \pi  \ell^{2}} -\frac{Q^2}{32 \pi  r^2} \label{T11q}\\
T^2_2&=&\frac{1}{8 \pi  \ell^{2}} +\frac{Q^2}{32 \pi  r^2}.\label{T22q}
\end{eqnarray}
Upon comparing (\ref{T00}-\ref{T22}) with equations (\ref{T00q}-\ref{T22q}), we establish the identification $Q^2=2M(\gamma-1)/(\eta-1)$. We can conclude, then, that the non-locality introduced by the fractional derivative leaves traces in the classical results, providing anisotropy to the cosmological constant and an effective electric charge. This result resembles, to some extent, the Kaluza-Klein mechanism (see \cite{Overduin:1997sri}, for example), where a gravitational theory in 5-dimensional spacetime leads to a theory of gravity coupled with electrodynamics in 4-dimensional spacetime after compactification. In this case, compactification is replaced by non-locality. We want to emphasize that the mechanism described here is unrelated to Kaluza-Klein, but the resemblance is interesting.
 \\

{Before concluding this section, we would like to make it clear that the primary reason for working in 2+1 dimensions is to make the problem more tractable from a computational and analytical standpoint. This dimensional reduction is not related to any fundamental aspect of our proposed fractional derivative but is instead a necessary step to address the technical challenges posed by fractional Einstein equations in higher dimensions. Specifically, fractional derivatives, such as the Riemann-Liouville operator, involve integro-differential terms that grow increasingly complex with higher-dimensional metrics. By restricting to a static and circularly symmetric 2+1-dimensional metric, the number of non-trivial field equations is reduced, allowing us to isolate the effects of the fractional derivative without overwhelming computational difficulties.
We acknowledge that the scope of the paper is narrow and focused on extracting specific results within a simplified framework. However, this is a deliberate first step toward understanding the implications of fractional calculus in General Relativity. The results obtained here provide a proof of concept for the feasibility of formulating fractional Einstein equations and exploring their consequences. Extending the framework to higher-dimensional spacetimes or more complex geometries (e.g., cosmological models or quantum gravitational settings) will undoubtedly introduce significant challenges, including increased computational complexity and new conceptual questions, such as non-local effects in evolving spacetimes.
While our work focuses on a specific and simplified scenario, the results are significant for several reasons. They demonstrate how fractional derivatives modify key features of classical solutions, such as the introduction of an anisotropic cosmological constant and effective charge. Additionally, they highlight the role of non-locality in extending General Relativity, opening the door for future studies to explore cosmological and quantum gravitational settings.
}



\section{conclusions}
{
The application of fractional calculus in General Relativity presents both theoretical and mathematical challenges, making it a compelling area of investigation due to its potential to address fundamental issues in gravitational theory. Moreover, with the increasing precision of astronomical and cosmological observations, the need to explore extensions beyond General Relativity has become more relevant than ever. With this motivation, our work explores the feasibility of formulating fractional Einstein field equations using a specific metric parameterization, akin to conventional differential operators.
\\
We introduced a novel fractional derivative that resolves key limitations of existing operators such as Riemann-Liouville and Caputo. Specifically, the Riemann-Liouville derivative does not vanish on constants, leading to spurious degrees of freedom, while the Caputo derivative, although resolving this issue, introduces divergences when constructing the Christoffel symbols. Our proposed weighted Riemann-Liouville derivative ensures that the derivative of a constant is zero, preventing unwanted coordinate dependencies, while also avoiding the divergences that arise with the Caputo formulation. 
\\
To apply this framework, we examined the general metric of a static and circularly symmetric 2+1-dimensional spacetime and derived a set of integro-differential equations. This choice of dimensionality was motivated by the need to keep the problem computationally and analytically tractable. Higher-dimensional formulations introduce increasingly complex integro-differential terms, which make direct solutions impractical. By restricting to a lower-dimensional setting, we were able to isolate the effects of fractional derivatives in a controlled manner while ensuring mathematical consistency.
\\
As a specific example, we analyzed the fractional version of the Bañados-Teitelboim-Zanelli (BTZ) metric, a vacuum solution of the Einstein field equations with a negative cosmological constant. Our findings indicate that, for fractional parameters close to one, the solution exhibits an anisotropic cosmological constant and an effective charge, similar to a charged BTZ solution but with an altered structure. This suggests that fractional modifications introduce non-local effects that can be interpreted as a form of Kaluza-Klein mechanism, potentially influencing the effective matter-energy distribution in gravitational systems. Exploring these effects in a cosmological setting could provide valuable insights and may offer testable predictions against observational data.
\\
Additionally, a key avenue for future research is the construction of a fully fractional metric that converges to the classical one in the appropriate limits. Since fractional derivatives inherently introduce non-locality, imposing a classical metric might be overly restrictive. Furthermore, from a mathematical perspective, it remains an open question whether a fractional Jacobi identity can be formulated to derive the Bianchi identities in a fractional setting. Another crucial step is the development of systematic methods for solving the integro-differential equations derived in this work, instead of assuming specific metric forms. Finally, a deeper exploration of how fractional derivatives affect the matter sector would be essential to fully understand the physical implications of this approach. 
\\
Despite the narrow scope of this work, our findings establish a proof of concept for the feasibility of fractional Einstein equations and their potential consequences. Extending this framework to higher-dimensional spacetimes or evolving cosmological models poses significant challenges, both computational and conceptual. However, our results demonstrate that fractional derivatives offer a promising avenue for incorporating non-local effects into gravitational theories, which could ultimately contribute to a broader understanding of gravity beyond General Relativity.
}
\\

\section{acknowledgements}
E. C. is funded by the Beatriz Galindo contract BG23/00163 (Spain). E. C. acknowledge Generalitat Valenciana through PROMETEO PROJECT CIPROM/2022/13.

\section{appendix}

\subsection{Fractional Derivatives of \( x^n \) in Different Definitions}

We explicitly compute the fractional derivatives of the power function \( x^n \) using the Riemann-Liouville, Caputo, and modified fractional derivative definitions.

\subsection*{1. Riemann-Liouville Fractional Derivative}
The left-sided Riemann-Liouville fractional derivative with $0<\eta<1$ is defined as:
\begin{equation}
D_{a+}^{\eta} x^n = \frac{1}{\Gamma(1-\eta)} \frac{d}{dx} \int_a^x (x-t)^{-\eta} t^n dt.
\end{equation}
Solving the integral by substitution and beta function:
\begin{equation}
I_{a+}^{1-\eta} x^n = \frac{ \Gamma(n+1)}{\Gamma(n+2-\eta)}x^{n+1-\eta}.
\end{equation}
Applying the derivative:
\begin{equation}
D_{a+}^{\eta} x^n = \frac{\Gamma(n+1)}{\Gamma(n+1-\eta)} x^{n-\eta}.
\end{equation}

\subsection*{2. Caputo Fractional Derivative}
The Caputo derivative is defined as:
\begin{equation}
D_{a+}^{\eta} x^n = \frac{1}{\Gamma(1-\eta)} \int_a^x (x-t)^{-\eta} \frac{d}{dt} (t^n) dt.
\end{equation}
Computing the derivative inside the integral:
\begin{equation}
D_{a+}^{\eta} x^n = \frac{n}{\Gamma(1-\eta)} \int_a^x (x-t)^{-\eta} t^{n-1} dt.
\end{equation}
Using the fractional integral identity:
\begin{equation}
D_{a+}^{\eta} x^n = \frac{n \Gamma(n)}{\Gamma(n+1-\eta)} x^{n-\eta}.
\end{equation}

\subsection*{3. Modified Fractional Derivative}
Using the modified definition:
\begin{equation}
(qD^{\eta}_{a+} x^n) = q_1(x, \eta) \frac{d}{dx} \left[ q_2(x, \eta) I_{a+}^{1-\eta} x^n \right].
\end{equation}
Computing the fractional integral:
\begin{equation}
I_{a+}^{1-\eta} x^n = \frac{\Gamma(n+1)}{\Gamma(n+2-\eta)}x^{n+1-\eta}.
\end{equation}
Applying the weighted derivative:
\begin{equation}
(qD^{\eta}_{a+} x^n) = q_1(x, \eta) \frac{d}{dx} \left[ q_2(x, \eta) \frac{\Gamma(n+1)}{\Gamma(n+2-\eta)}x^{n+1-\eta} \right].
\end{equation}
Expanding the derivative:
\begin{equation}
(qD^{\eta}_{a+} x^n) = q_1(x, \eta) \left[ q_2(x, \eta) \frac{\Gamma(n+1)x^{n-\eta}}{\Gamma(n+1-\eta)} + \frac{dq_2}{dx} \frac{\Gamma(n+1)x^{n+1-\eta}}{\Gamma(n+2-\eta)} \right].
\end{equation}

This formulation shows that the modified derivative introduces additional correction terms via the weight functions \( q_1(x, \eta) \) and \( q_2(x, \eta) \), allowing for tunable behavior that can be adjusted to avoid issues found in Riemann-Liouville and Caputo derivatives.

\end{document}